\documentclass[]{elsarticle}

\usepackage{lineno,hyperref}
\modulolinenumbers[5]

\journal{Journal of Physica C: Superconductivity and its Applications}

\begin{document}

\def\rr{{{\bf r}}}
\def\rrp{{{\bf r}^\prime}}
\def\ll{{{\textbf{\textit l}}}}
\def\llp{{{\textbf{\textit l}}^\prime}}

\begin{frontmatter}

\title{Electronic Structure Properties of UO2 as a Mott Insulator}

\author[]{Samira Sheykhi}

\author[]{Mahmoud Payami\corref{mycorrespondingauthor}}
\cortext[mycorrespondingauthor]{Corresponding author}
\ead{mpayami@aeoi.org.ir}


\address[]{Physics \& Accelerators Research School, Nuclear Science and Technology Research Institute, P. O. Box 14395-836, Tehran, Iran}

\begin{abstract}
In this work using the density functional theory (DFT), we have studied the structural, electronic and magnetic properties of uranium dioxide with antiferromagnetic 1k-, 2k-, and 3k-order structures.  
Ordinary approximations in DFT, such as the local
density approximation (LDA) or generalized gradient approximation (GGA), usually predict incorrect metallic behaviors for this strongly correlated electron system. Using Hubbard term correction for f-electrons, LDA+U method, as well as using the screened Heyd-Scuseria-Ernzerhof (HSE) hybrid functional for the exchange-correlation (XC), we have obtained the
correct ground-state behavior as an insulator, with band gaps in good agreement with experiment. 
\end{abstract}

\begin{keyword}
Mott insulator; Antiferromagnetic; Uranium dioxide; Density functional theory; Hybrid functional
\end{keyword}

\end{frontmatter}


\section{\label{sec1}Introduction}
\begin{figure}
\begin{center}
\includegraphics[width=0.7\linewidth]{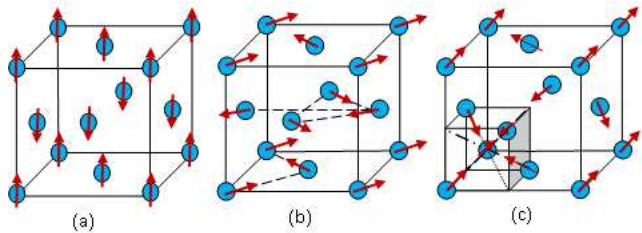}
\caption{\label{fig1} Uo$_2$ in (a) 1k-order, (b) 2k-order, and (c) 3k-order antiferromagnetic structures. }
\end{center}
\end{figure}

Urania (UO2) is one of the most common fuels used in nuclear power reactors, and understanding its properties, both experimentally \cite{wilkins2006direct} and theoretically \cite{freyss4,laskowski2004magnetic,peng2015electronic,allen2014occupation}, has been one of the hot topics of research activity for several years.
The crystal structure of urania has been experimentally determined to be an antiferromagnetic 3k-order Mott
insulator at temperatures below 30~$^\circ$K, in which the uranium atoms are located in an fcc lattice with a=b=c=5.47$\AA$, and the oxygen atoms are located at positions with  symmetry $Pa\bar{3}$ \cite{idiri2004behavior}. In Fig~\ref{fig1}(a)-(c), we have shown the antiferromagnetic 1k-, 2k-, and 3k-order structures. 

\begin{figure}
\begin{center}
\includegraphics[width=0.8\linewidth]{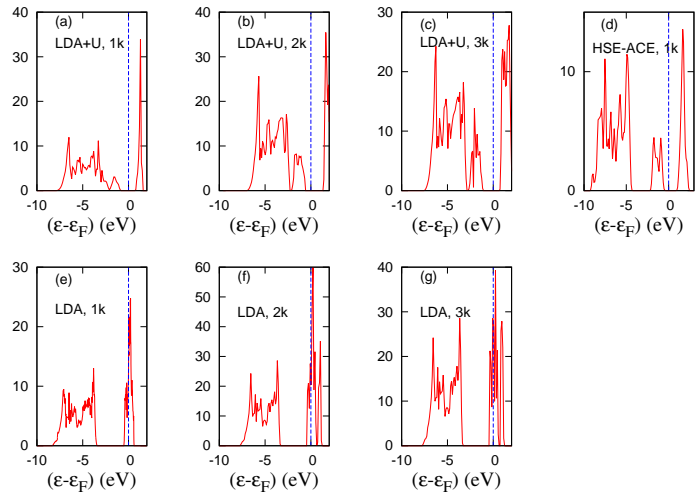}
\caption{\label{fig2} Density of states: (a)-(c) for LDA+U, (d) for HSE, (e)-(g) for LDA. The zero of energy is identified as the Fermi level. }
\end{center}
\end{figure}

DFT is one of the most powerful methods for calculation of electronic structure properties of materials \cite{hohenberg1964inhomogeneous,kohn1965self}. Application of ordinary DFT, using LDA or GGA for the exchange-correlation functional, incorrectly predicts a metal behavior (see Fig~\ref{fig2}) for urania, while it is known as a Mott insulator solid. This behavior is due to the localized ``d'' and ``f'' electrons that in ordinary DFT calculations are accounted as nearly delocalized ones which is appropriate for s-p bonded solid materials. One of the methods to rectify this deficiency is the use of DFT+U method \cite{liechtenstein1995density} in which a Hubbard term for the localized ``f'' electrons is added to the Hamiltonian, and this correction leads to an opening of a gap, leading to an insulating properties. The method is somewhat tricky and to obtain the true ground state, one has to avoid metastable states \cite{freyss4}. Another method, which is rather new and more appropriate, is using the screened HSE hybrid functional \cite{heyd2003hybrid} which partially uses the exact exchange operator. Using this orbital-dependent functional is computationally very expensive and the newly formulated Adaptively Compressed Exchange (ACE) Operator \cite{lin2016adaptively} significantly reduces the computational efforts while keeping its accuracy. In this work, we have used LDA and LDA+U methods for 1k, 2k, and 3k antiferromagnetic configurations. The HSE-ACE method is applied only to the 1k-order. All LDA+U results as well as the HSE-ACE, correctly predict insulating behaviors for UO$_2$. 

\section{Calculation methods}\label{sec2}
All calculations are based on DFT and the solution of the Kohn-Sham
equations \cite{kohn1965self} using the Quantum-ESPRESSO code package \cite{giannozzi2009quantum}. For the atoms U
and O, we have used the pseudopotentials U.pz-mt-fhi.UPF and O.pz-mt-fhi.UPF which are
available at http://www.quantum-espresso.org. By convergency tests, the kinetic energy cutoffs for the plane-wave
bases are chosen as 40 and 160 Ry for the wave functions and charge density, respectively. In HSE calculations, the plane-wave cutoff for expansion of the Fock exchange operator is taken as 80~Ry. For
the Brillouin-zone integrations, a $4\times4\times4$ grid with a shift were used. In LDA+U calculations, we
have used the values 4.5 and 0.54 eV for U and J parameters, respectively as determined by other
works \cite{yamazaki1991systematic,kotani1992systematic}. All geometries are fully optimized for pressures on unit cells to within 1kPa, and forces on atoms to within 1mRy/a.u.
\section{Results and Discussions}
The LDA calculations for all UO$_2$ antiferromagnetic structures lead to metallic states, as are shown by density-of-states (DOS) plots in Figs.~\ref{fig2}(e)-(g).   
The LDA+U calculations for the 1k-order configuration, led to a small
deformation in c-axis corresponding to a tetragonal structure with a=b=5.49, and c=5.44\AA , with
a gap of 1.5~eV.

As to 2k-order configuration, locating the ground-state in LDA+U was somewhat tricky in that one should avoid metastable states, and the true ground state was achieved by some modifications in the occupation matrix. The results show a similar deformation as in 1k-order but with a=b=5.50, and c=5.49\AA, with 1.9~eV gap.

For 3k-order, the optimization of the structure maintained the cubic structure with a=b=c=5.50\AA, but with a small Jahn-Teller distortion for the oxygen atoms (0.0153\AA in $<111>$ direction) in good agreement with experiment. The energy gap was found to be 1.7~eV, in good agreement with the experiment (2.1~eV). Here, locating the true ground state was also tricky as in 2k case. 

Finally, the HSE-ACE calculations for the 1k-order led to a similar tetragonal structure with a=b=5.53, and c=5.54\AA. In this calculation, the c-axis is slightly elongated in contrast to our LDA+U results. It has been shown that the $c<a$ or $c>a$ behavior depends on the choice of Hubbard U and J values \cite{yu2009first}. This calculation relulted in a gap of 2.02~eV which is quite in good agreement with experiment as compared with the LDA+U results. The DOS plots for LDA+U and HSE-ACE are shown in Figs.~\ref{fig2}(a)-(d), respectively. It should be mentioned that the lower band-gap shown in Fig~\ref{fig2}(d) compared to those of LDA+U (whereas it should be higher) is due to low-density k-mesh for producing the DOS-plot. We have presented the HSE-ACE plot only to show that HSE correctly reproduce the insulating behavior; the band gap is simply read off from the calculated Kohn-Sham eigenvalues.       

\section{\label{sec5}Conclusions}
LDA or GGA calculations for UO$_2$ lead to an incorrect metallic behavior. This is corrected using the LDA+U method and using HSE hybrid functional. The 1k- and 2k-order antiferromagnetic structures lead to a small deformation in the $c$-axis, while the 3k-order LDA+U preserves the cubic symmetry for U atoms even though the O atoms are slightly distorted along the $<111>$ direction. All LDA+U and HSE calculations led to correct insulating behaviors. More interestingly, the gap reproduced by HSE calculation is in a very good agreement with experimental value.
\section*{Acknowledgements}
This work is part of research program in Theoretical and Computational Physics Group, NSTRI, AEOI.
\section*{References}
\bibliography{payami-physicaci-13July17} 
\end{document}